\documentclass[runningheads]{llncs}
\def\BibTeX{{\rm B\kern-.05em{\sc i\kern-.025em b}\kern-.08em
   T\kern-.1667em\lower.7ex\hbox{E}\kern-.125emX}}
\usepackage{graphicx}
\usepackage{mathtools}
\usepackage[utf8]{inputenc}
\usepackage{lipsum} 
\usepackage{booktabs}
\usepackage[ruled]{algorithm2e}
\usepackage{url}
\usepackage{colortbl}
\usepackage{multicol}
\usepackage{color}
\usepackage{multirow}
\usepackage{enumitem}
\usepackage{array}
\usepackage{siunitx}
\usepackage[labelfont={bf}]{caption}
\usepackage{balance}
\usepackage{graphicx}
\usepackage{array} 
\usepackage{todonotes}
\usepackage{subfig}
\usepackage[nolist,nohyperlinks]{acronym}
\usepackage{footnote}
\usepackage[perpage]{footmisc}
\usepackage[autostyle]{csquotes}
\makesavenoteenv{tabular}
\acrodef{DL}{Deep learning}
\acrodef{CRC}{colorectal cancer}
\acrodef{CNN}{Convolutional Neural Network}
\acrodef{CADx}{computer aided diagnosis} 
\acrodef{SOTA}{state-of-the-art}
\acrodef{mIoU}{mean intersection over Union}
\acrodef{FPS}{frame per second}
\acrodef{AI}{Artificial Intelligence}
\acrodef{ML}{Machine Learning}
\acrodef{ROI}{Region of Interest}
\acrodef{DSC}{dice coefficient}
\begin{document}

\title{DilatedSegNet: A Deep Dilated Segmentation Network for Polyp Segmentation}
\titlerunning{DilatedSegNet: A Deep Dilated Segmentation Network for Polyp Segmentation}
\author{
  Nikhil Kumar Tomar, Debesh Jha, Ulas Bagci}
\authorrunning{N. Tomar et al.}
\institute{Machine \& Hybrid Intelligence Lab, Department of Radiology,\\ Northwestern University}

\maketitle   
\begin{abstract}
Colorectal cancer (CRC) is the second leading cause of cancer-related death worldwide. Excision of polyps during colonoscopy helps reduce mortality and morbidity for CRC. Powered by deep learning, computer-aided diagnosis (CAD) systems can detect regions in the colon overlooked by physicians during colonoscopy. Lacking high accuracy and real-time speed are the essential obstacles to be overcome for successful clinical integration of such systems. While literature is focused on improving accuracy, the speed parameter is often ignored. Toward this critical need, we intend to develop a novel real-time deep learning-based architecture, DilatedSegNet, to perform polyp segmentation on the fly. DilatedSegNet is an encoder-decoder network that uses pre-trained ResNet50 as the encoder from which we extract four levels of feature maps. Each of these feature maps is passed through a dilated convolution pooling (DCP) block.  The outputs from the DCP blocks are concatenated and passed through a series of four decoder blocks that predicts the segmentation mask. The proposed method achieves a real-time operation speed of 33.68 frames per second with an average \ac{DSC} of 0.90 and mIoU of 0.83. Additionally, we also provide heatmap along with the qualitative results that shows the explanation for the polyp location, which increases the trustworthiness of the method. The results on the publicly available Kvasir-SEG and BKAI-IGH datasets suggest that DilatedSegNet can give real-time feedback while retaining a high \ac{DSC}, indicating high potential for using such models in real clinical settings in the near future. The GitHub link of the source code can be found here: \url{https://github.com/nikhilroxtomar/DilatedSegNet}.

\end{abstract}

\section{Introduction}
Missed polyp during routine colonoscopy examination is the primary source of interval \acf{CRC}. The polyps that are not recognized within the colonoscope are the major source contributor to this problem. Colonoscopy is considered the gold standard for colon cancer diagnosis and follow-up.  However, 22-28\% of polyps are missed during a routine examination~\cite{leufkens2012factors}. Some of these polyps can cause post-colonoscopy colorectal cancer (CRC). One of the reasons for the polyp miss-rate is either the polyp was not visible during the examination or was not recognized despite being in the visual field because of the faster colonoscope withdrawal time. Deep learning based algorithms can highlight the presence of pre-cancerous tissue in the colon and have the potential to improve the diagnostic performance of endoscopists. Improving the polyp detection rate as well as its accurate segmentation is an unmet clinical need. In practice, precise polyp segmentation provides important information in the early detection of colorectal cancer via their shape, texture, and location information.

Tomar et al.~\cite{tomar2022fanet} proposed a feedback attention network for biomedical image segmentation where they utilized the previous epoch mask with the current training epoch in an iterative fashion to further improve the performance. Fan et al.~\cite{fan2020pranet} used Res2Net-based~\cite{gao2019res2net} backbone where they used a parallel partial decoder and parallel reverse attention mechanism for the accurate polyp segmentation. Jha et al.~\cite{jha2019resunet++} proposed an efficient architecture where they utilized the strength of the residual block, atrous spatial pyramidal pooling, with squeeze and excitation block for polyp segmentation. Shen et al.~\cite{shen2021hrenet} proposed a hard region enhancement network (HRENet) that consists of an informative context enhancement (ICE) module and trained the model on edge and structure consistency aware loss (ESCLoss) to improve the polyp segmentation on the precise edge. Zhao et al.~\cite{zhao2021automatic} proposed a multi-scale subtraction network (MSNet) for automatic polyp segmentation. Despite of several architectures proposed in the literature, most existing methods often neglect the encoder and tend to focus more on the decoder part of the network, which led to the loss of significant features from the encoder part. In our proposed method, we focus more on the encoder part of the network by utilizing different scales features which are passed through multiple dilated convolutions to capture more enlarged features, leading to improved polyp segmentation. Unlike other decoders, the design of our decoder is straightforward. It utilizes simple sequences of layers such as an upsampling layer, concatenation, residual block and an attention layer. We introduce the novel deep learning architecture, DilatedSegNet, to address the critical need for clinical integration of polyp segmentation routine, which is real-time and retains high accuracy. The main contribution of the study are as follows:
\begin{enumerate}
 \item We introduce a novel network named DilatedSegNet for polyp segmentation. The architecture begins with a pre-trained ResNet50~\cite{he2016deep} and utilizes dilated convolution~\cite{yu2015multi} pooling block to increase the receptive field for capturing more diverse and reliable features for a better delineation. 
 
    \item DilatedSegNet showed outstanding performance by outperforming nine standard benchmarking methods with two widely used publicly available polyp segmentation datasets. 
    
    \item Extensive experimental results and cross-dataset test results on two unseen datasets showed the better generalizability capability of the DilateSegNet. Explored deep features showed via heatmaps that the proposed network model is focusing on the target polyp regions and their boundaries, proving visual interpretability of the model.
\end{enumerate}

\section{Method}

\begin{figure*}[!t]
    \centering
    \includegraphics[width=0.999\textwidth]{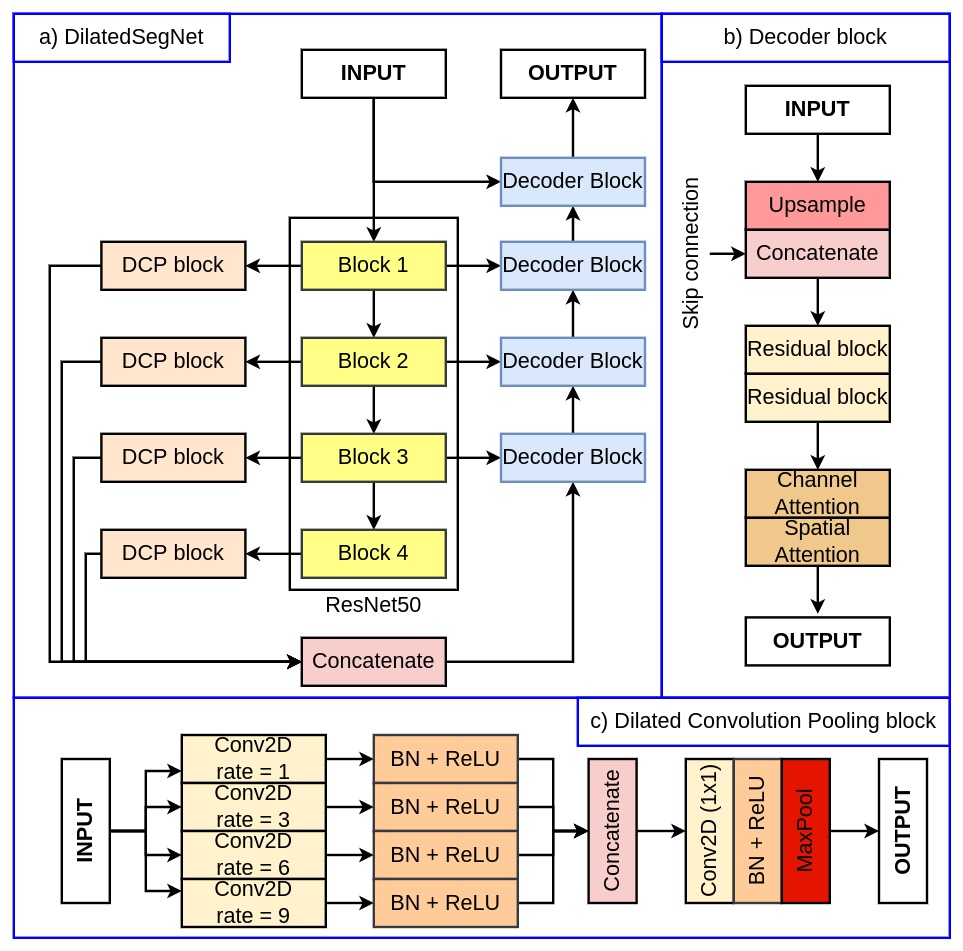}
    \caption{Block diagram of the proposed DilatedSegNet along with its components.}
    \label{fig:architecture}
\end{figure*}
Figure~\ref{fig:architecture} shows the block diagram of the proposed DilatedSegNet along with its core components. It follows an encoder-decoder scheme much like the U-Net~\cite{ronneberger2015u}, consisting of a pre-trained ResNet50~\cite{he2016deep} as an encoder. The input image $I$ with a resolution of [$h\times w\times3$] is fed to the pre-trained encoder from which we extracts four levels of features maps \{$f_{i}: i=1,2,3,4$\} with varying resolution of [$h/2^k \times w/2^k: k=1,2,3,4$]. Each of these feature maps is then passed through a Dilated Convolution Pooling (DCP) block, where four parallel dilated convolutions with the rate $1, 3, 6, 9$ are applied to enhance the field of view. The output from all the DCP blocks is concatenated and passed to the first decoder block, where the feature map is upsampled and concatenated with a skip connection from the pre-trained encoder. Next, it is passed through some residual block and then a Convolutional Block Attention Module (CBAM)~\cite{woo2018cbam}. The output of the CBAM is passed to the next decoder for further transformation. Finally, the output from the last decoder block is passed to a $1\times1$ convolution followed by a sigmoid activation function.

\subsection{Dilated Convolution Pooling (DCP) block}
The DCP block begins with four parallel $3\times3$ convolution layers having a dilation rate of $1$, $3$, $6$ and $9$. The dilated convolution increases the receptive field of the $3\times3$ kernel, which helps it to cover more area over the input feature maps. Thus, by increasing dilation rate, we get better feature maps with each layer. The output from each convolutional layer is followed by batch normalization and a ReLU activation function. Next, we combine the output from each ReLU activation function to form a feature map, which is followed by a $1\times1$ convolutional layer to reduce the number of feature channels. The $1\times1$ convolutional layer is further followed by batch normalization and a ReLU activation function. The output of the ReLU activation function is passed through a max-pooling layer to reduce its spatial dimensions.

\subsection{Decoder block}
The decoder block begins with a bilinear upsampling where the feature map spatial dimensions (height and width) are increased by a factor of two. After that, we concatenate the upsampled feature map with the feature map from the pre-trained encoder through the skip connections. These skip connections fetch the necessary features directly from the encoder to the decoder which is sometimes lost due to the depth of the network. The concatenated feature maps are passed through a set of two residual blocks which helps to learn more meaningful semantic features from the input. These features are further refined by using an attention mechanism called CBAM\cite{woo2018cbam}. CBAM consists of channel attention followed by spatial attention to highlight the more significant features and suppress the irrelevant ones. 

\section{Experimental setup}

\begin{table*}[t!]
\centering
\caption{Complexity of the models with image size of $256\times 256$.}

\begin{tabular} {@{}l|l|c|c|c|c@{}}
\toprule
\textbf{Method} & \textbf{\shortstack{Publication\\ Venue}} & \textbf{Backbone} & \textbf{\shortstack{Parameters \\(Millions)}} & \textbf{\shortstack{Flops\\(GMac)}}  & \textbf{FPS} \\
\hline

U-Net~\cite{ronneberger2015u} & MICCAI'15 & - &31.04 &54.75 & 160.27 \\
ResU-Net~\cite{zhang2018road} & GRSL'18 & - &8.22 &45.42 & 197.94\\
U-Net++~\cite{zhou2018unet++} &DLMIAW'18 & - &9.16 &34.65 & 123.45 \\
DeepLabV3+~\cite{chen2018encoder} &ECCV'18 & ResNet50 &39.76 &43.31 & 99.16 \\
ResU-Net++~\cite{jha2019resunet++} &ISM'19 & - &4.06 &15.81 & 55.86 \\
DDANet~\cite{tomar2021ddanet} &ICPRW'20 & - &3.36 &18.2 & 86.46 \\
PraNet~\cite{fan2020pranet} &MICCAI'20 & Res2Net &32.55 &6.93 &36.21 \\
ColonSegNet~\cite{jha2021real} &IEEE Access'21 & - &5.01 &62.16 & 122.42\\
HarDNet-MSEG~\cite{huang2021hardnet} &Arxiv'21 & - &33.34 &6.02 & 41.20\\
FANet~\cite{tomar2022fanet} &IEEE TNNLS'22 & - &7.72 &94.75 &65.53 \\
CaraNet~\cite{lou2022caranet} &SPIE MI'22 & Res2Net & 46.64 &11.48 &20.13 \\
\textbf{DilatedSegNet (Ours)} & - & ResNet50 &18.11 &27.1 & 33.68 \\
\hline
\end{tabular}
\label{tab:complexity}
\end{table*}

In this section, we present the datasets, evaluation metrics and the implementation details.

\subsection{Datasets and evaluation metrics}
We have selected the Kvasir-SEG~\cite{jha2020kvasir} and BKAI-IGH~\cite{lan2021neounet} datasets to evaluate the performance of the proposed DilatedSegNet. Both of the datasets are publicly available and can be easily accessible. Kvasir-SEG~\cite{jha2020kvasir} consists of 1000 polyp images, their corresponding masks, and the bounding box information. Similarly, BKAI-IGH~\cite{lan2021neounet} consists of 1000 polyp images in the training dataset, and separate 200 images in the test dataset. However, the ground truth of the test dataset is not made publicly available by the dataset provided. So, we only experiment on the training dataset. Additionally, this dataset consists of neo-plastic and non-neoplastic. However, we treat the dataset as a binary class problem. We have used standard segmentation metrics such as \ac{DSC}, \ac{mIoU}, precision, recall, F2-score and \ac{FPS} to benchmark the performance of our proposed model.

\subsection{Implementation details}
In this study, we have implemented our proposed DilatedSegNet and all the other benchmark models using the PyTorch framework and trained on a RTX 3090 GPU. We have used the same hyperparameters for all the models for a fair comparison. The images and masks were first split into training, validation and testing datasets. For Kvasir-SEG, we have utilized the official split of $880/120$, where $880$ images and masks were used for training and the rest of the $120$ were used for validation and testing. For the BKAI dataset, we followed an $80:10:10$ split, where $80\%$ images and masks were used for training, $10\%$ was used for validation and the remaining $10\%$ was used for the testing. All the images and masks were resized to $256 \times 256$ pixels. To make the model more robust, we have used an online data augmentation strategy with random rotation, horizontal flipping, vertical flipping and coarse dropout. All the models were trained by an Adam optimizer~\cite{kingma2014adam} with a learning rate of $1e-4$ and a batch size of $16$. We have used a combination of dice loss and binary cross-entropy as the loss function. ReduceLROnPlateau was used while training to reduce the learning rate for better performance, while early stopping was used to stop the training when the model stopped improving.

\begin{table*}[t!]
\centering
\caption{Results on the Kvasir-SEG~\cite{jha2020kvasir} and BKAI-IGH~\cite{lan2021neounet} datasets.}
 \begin{tabular} {@{}l|l|c|c|c|c|c@{}}
\toprule
\textbf{Method} &\textbf{Publication} &\textbf{DSC} & \textbf{mIoU} &\textbf{Recall}  &\textbf{Precision} &\textbf{F2}\\ 
\hline
\multicolumn{6}{@{}l}{\textbf{Train and test data: Kvasir-SEG~\cite{jha2020kvasir}}}  \\ \hline
U-Net~\cite{ronneberger2015u} &MICCAI'15	&0.8264	&0.7472	&0.8504	&0.8703	&0.8353	\\
ResU-Net~\cite{zhang2018road} &GRSL'18 &	0.7642	&0.6634	&0.8025	&0.8200	&0.7740	\\
U-Net++~\cite{zhou2018unet++} &DLMIAW'18 &0.8228	&0.7419	&0.8437	&0.8607	&0.8295	\\
DeepLabV3+~\cite{chen2018encoder} &ECCV'18 &0.8837	&0.8173	&0.9014	&0.9028	&0.8904\\
ResU-Net++~\cite{jha2019resunet++} &ISM'19&0.6453	&0.5341	&0.6964	&0.7080	&0.6575\\
DDANet~\cite{tomar2021ddanet} &ICPRW'20 &0.7415	&0.6448	&0.7953	&0.7670	&0.7640\\
PraNet~\cite{fan2020pranet}  &MICCAI'20& 0.8942&0.8296&	0.9060&	\textbf{0.9126}&0.8976 \\
ColonSegNet~\cite{jha2021real} &IEEE Access'21 &0.7920	&0.6980	&0.8193	&0.8432	&0.7999\\
HarDNet-MSEG~\cite{huang2021hardnet} &Arxiv'21 &0.8260	&0.7459	&0.8485	&0.8652	&0.8358\\
FANet~\cite{tomar2022fanet} &IEEE TNNLS’22 &0.7844 &0.6975 &0.8503 &0.8165 &0.8054 \\
CaraNet~\cite{lou2022caranet} & SPIE MI'22 &0.8742 &0.8001 &\textbf{0.9289} &0.8614 &0.8996 \\
\textbf{DilatedSegNet (Ours)} & - & \textbf{0.8957}& \textbf{0.8336}& 0.9169& 0.9096& \textbf{0.9034}\\

\midrule
\multicolumn{6}{@{}l}{\textbf{Train and test data: BKAI-IGH~\cite{lan2021neounet}}}  \\ \hline
U-Net~\cite{ronneberger2015u} &MICCAI'15 	&0.8286	&0.7599	&0.8295	&0.8999	&0.8264	\\
ResU-Net~\cite{zhang2018road} &GRSL'18 	&0.7433	&0.6580	&0.7447	&0.8711	&0.7387	\\
U-Net++~\cite{zhou2018unet++}& DLMIAW'18&0.8275	&0.7563	&0.8388	&0.8942	&0.8308	\\
DeepLabV3+~\cite{chen2018encoder} & ECCV'18 &0.8937	&0.8314	&0.8870	&\textbf{0.9333}&0.8882	\\
ResU-Net++~\cite{jha2019resunet++}& ISM'19  &0.7130	&0.6280	&0.7240	&0.8578	&0.7132\\
DDANet~\cite{tomar2021ddanet}& ICPRW'20	&0.7269	&0.6507	&0.7454	&0.7575	&0.7335\\
PraNet~\cite{fan2020pranet} &MICCAI'20 &0.8904	&0.8264	&0.8901 &0.9247 & 0.8885\\
ColonSegNet~\cite{jha2021real} &IEEE Access'21 	&0.7748	&0.6881	&0.7852	&0.8711	&0.7746\\
HarDNet-MSEG~\cite{huang2021hardnet} &Arxiv'21	&0.7627	&0.6734	&0.7532	&0.8344	&0.7528\\
FANet~\cite{tomar2022fanet} &IEEE TNNLS’22 &0.8305 &0.7578 &0.8285 &0.9169 &0.8243 \\
CaraNet~\cite{lou2022caranet} &SPIE MI'22 &0.8948 &0.8309 &0.8907 &0.9280 &0.8911 \\
\textbf{DilatedSegNet (Ours)} & - &\textbf{0.8950}& \textbf{0.8315}& \textbf{0.9082}& 0.9111& \textbf{0.8991}\\
\bottomrule
\end{tabular}
\label{tab:results-kvasir}
\end{table*}

\section{Results}
We present quantitative and qualitative results along with the heatmaps for model interpretability. 

\begin{figure*}[!ht]
    \centering
    \includegraphics[width=0.93\linewidth]{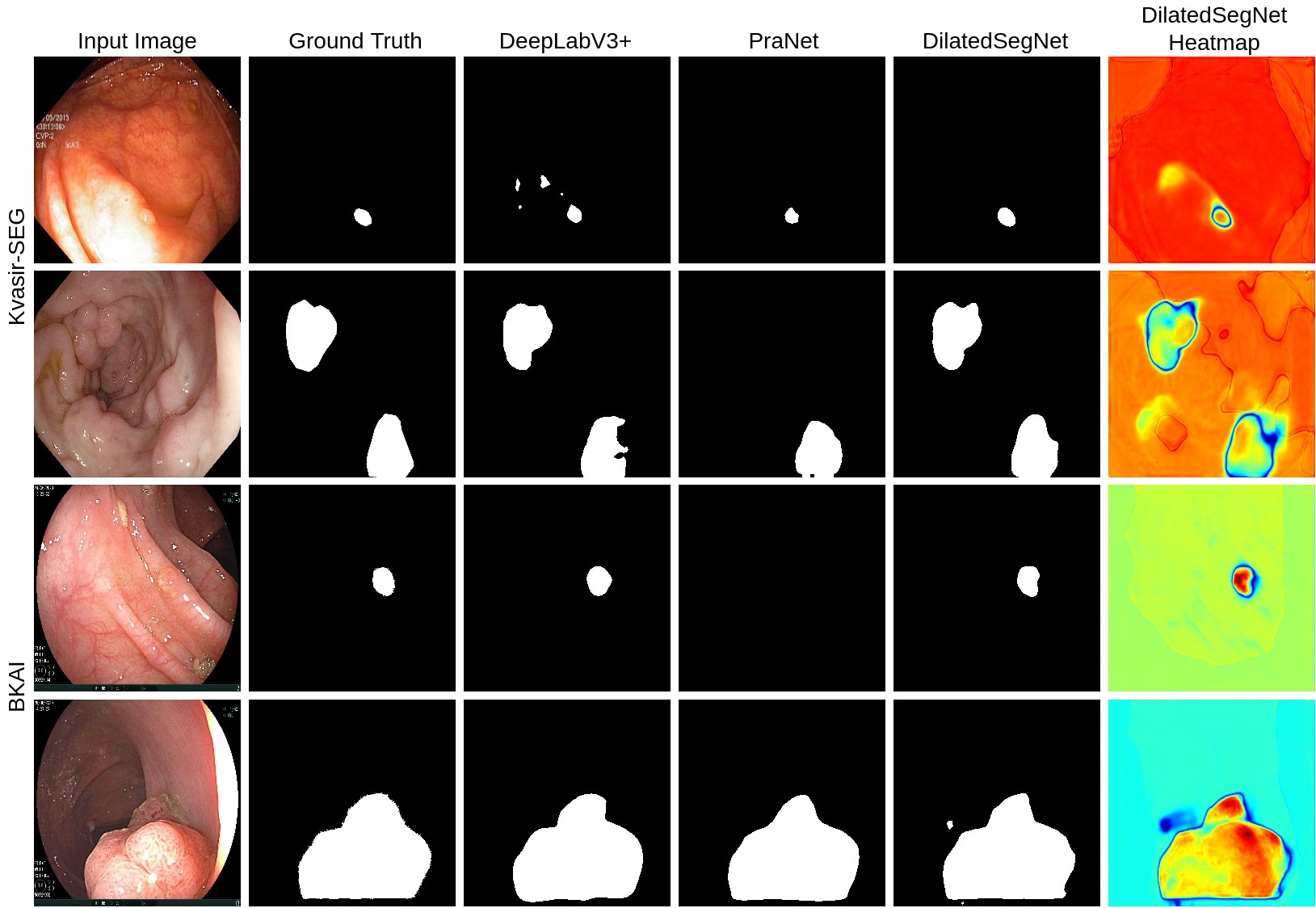}
    \caption{The figure shows qualitative results comparison of the three best methods. The heatmaps are obtained with respect to the convolutional layer at the bottleneck. The produced heatmap shows both important and unimportant pixels. Here, the heatmap shows that DilatedSegNet utilized correct pixels from the input image while making predictions for polyp and non-polyps. The qualitative comparison between the ground truth and the heatmap produced by DilatedSegNet shows that the heatmap is precise. This show that the prediction made by the proposed model is trustworthy.}
    \label{fig:qualitative}
\end{figure*}

\begin{table*}[t!]
\centering
\caption{Cross dataset results of models trained on Kvasir-SEG~\cite{jha2020kvasir} and tested on independent CVC-ClinicDB~\cite{bernal2015wm} and BKAI-IGH~\cite{lan2021neounet}.}
 \begin{tabular} {@{}l|l|c|c|c|c@{}}
\toprule
\textbf{Model} &	\textbf{DSC} &	\textbf{mIoU}&	\textbf{Recall}&	\textbf{Precision} &\textbf{F2}	\\
\midrule
\multicolumn{6}{@{}l}{\textbf{Train: Kvasir-SEG~\cite{jha2020kvasir}, Test: CVC-ClinicDB~\cite{bernal2015wm}}}  \\ \hline
U-Net~\cite{ronneberger2015u}&	0.6336	&0.5433	&0.6982	&0.7891	&0.6563	\\
ResU-Net~\cite{zhang2018road}	&0.5970	&0.4967	&0.6210	&0.8005 &0.5991	\\
U-Net++~\cite{zhou2018unet++}	&0.6350	&0.5475	&0.6933	&0.7967	&0.6556	\\
DeepLabV3+~\cite{chen2018encoder}	&0.8142	&0.7388	&0.8331	&0.8735	&0.8198	\\
ResU-Net++~\cite{jha2019resunet++}	&0.4642	&0.3585	&0.5880	&0.5770	&0.5084	\\
DDANet~\cite{tomar2021ddanet}	&0.5234	&0.4183	&0.6502	&0.5935	&0.5718\\
PraNet~\cite{fan2020pranet}	&0.8046	&0.7286	&0.8188	&\textbf{0.8968}	&0.8077\\
ColonSegNet~\cite{jha2021real}	&0.6126	&0.5090	&0.6564	&0.7521	&0.6246	\\
HarDNet-MSEG~\cite{huang2021hardnet}	&0.6960	&0.6058	&0.7173	&0.8528	&0.7010	\\
FANet~\cite{tomar2022fanet} &0.6524 &0.5579 &0.7560 &0.7243 &0.6872 \\
CaraNet~\cite{lou2022caranet} &0.8254 &0.7450 &\textbf{0.8568} &0.8696 &\textbf{0.8389} \\
\textbf{DilatedSegNet (Ours)}  &\textbf{0.8278}	&\textbf{0.7545} &0.8462 &0.8921 &0.8336\\
\midrule
\multicolumn{6}{@{}l}{\textbf{Train: Kvasir-SEG~\cite{jha2020kvasir}, Test: BKAI-IGH~\cite{lan2021neounet}}}  \\ \hline
U-Net~\cite{ronneberger2015u}&0.6347&0.5686	&0.6986	&0.7882	&0.6591\\
ResU-Net~\cite{zhang2018road}&0.5836&0.4931	&0.6716	&0.6549	&0.6177\\
U-Net++~\cite{zhou2018unet++}&0.6269&0.5592	&0.6900	&0.7968	&0.6493\\
DeepLabV3+~\cite{chen2018encoder}&0.7286 &0.6589	&0.7919	&0.8123	&0.7493\\
ResU-Net++~\cite{jha2019resunet++}&0.4166	&0.3204	&0.6979	&0.3922	&0.5019\\
DDANet~\cite{tomar2021ddanet}&0.5006&0.4115	&0.6612	&0.4825	&0.5592\\
PraNet~\cite{fan2020pranet}&0.7298	&0.6609	&0.8007 &0.824&0.7484 \\
ColonSegNet~\cite{jha2021real}&0.5765	&0.4910	&0.7191	&0.6644	&0.6225\\
HarDNet-MSEG~\cite{huang2021hardnet}&0.6502	&0.5711	&0.7420	&0.7469	&0.6830\\
FANet~\cite{tomar2022fanet} &0.5153 &0.4412 &\textbf{0.8395} &0.5505 &0.5913 \\
CaraNet~\cite{lou2022caranet} &0.7470 &0.6749 &0.8234 &0.8102 &\textbf{0.7742} \\
\textbf{DilatedSegNet (Ours)}&\textbf{0.7545} &\textbf{0.6906} &0.7886 &\textbf{0.8750}	&0.7649\\
\bottomrule
\end{tabular}
\label{tab:cross-data}
\end{table*}

\subsection{Performance test on same dataset}
Table~\ref{tab:results-kvasir} shows the result of the DilatedSegNet on the Kvasir-SEG~\cite{jha2020kvasir} and BKAI-IGH~\cite{lan2021neounet} datasets, respectively. DilatedSegNet obtains an \ac{DSC} score of 0.8957 and mIoU of 0.8336 with Kvasir-SEG and an \ac{DSC}-score of 0.8950 and mIoU of 0.8315 on the BKAI-IGH dataset, outperforming nine state-of-the-art benchmarks. The most competitive network to our network was PraNet~\cite{fan2020pranet} which obtained \ac{DSC} and mIoU 0.8942 and 0.8296, respectively, for the Kvasir-SEG. DeepLabv3+ obtained the most competitive results with BKAI-IGH: a \ac{DSC} of 0.8937 and mIoU of 0.8314. DilatedSegNet achieved a real-time operation speed of real-time speed of 33.68 FPS. The number of parameters used in DeepLabv3+ was 39.76 million and the number of flops utilized was 43.31 GMac. However, our proposed architecture has only 18.11 million parameters and 27.1 GMac flops (refer Table~\ref{tab:complexity}), substantially better performance by lowering the parameters and flops, thanks to our lightweight architectural design allowing for real-time processing.

\begin{figure*}[!ht]
    \centering
    \includegraphics[width=0.99\textwidth]{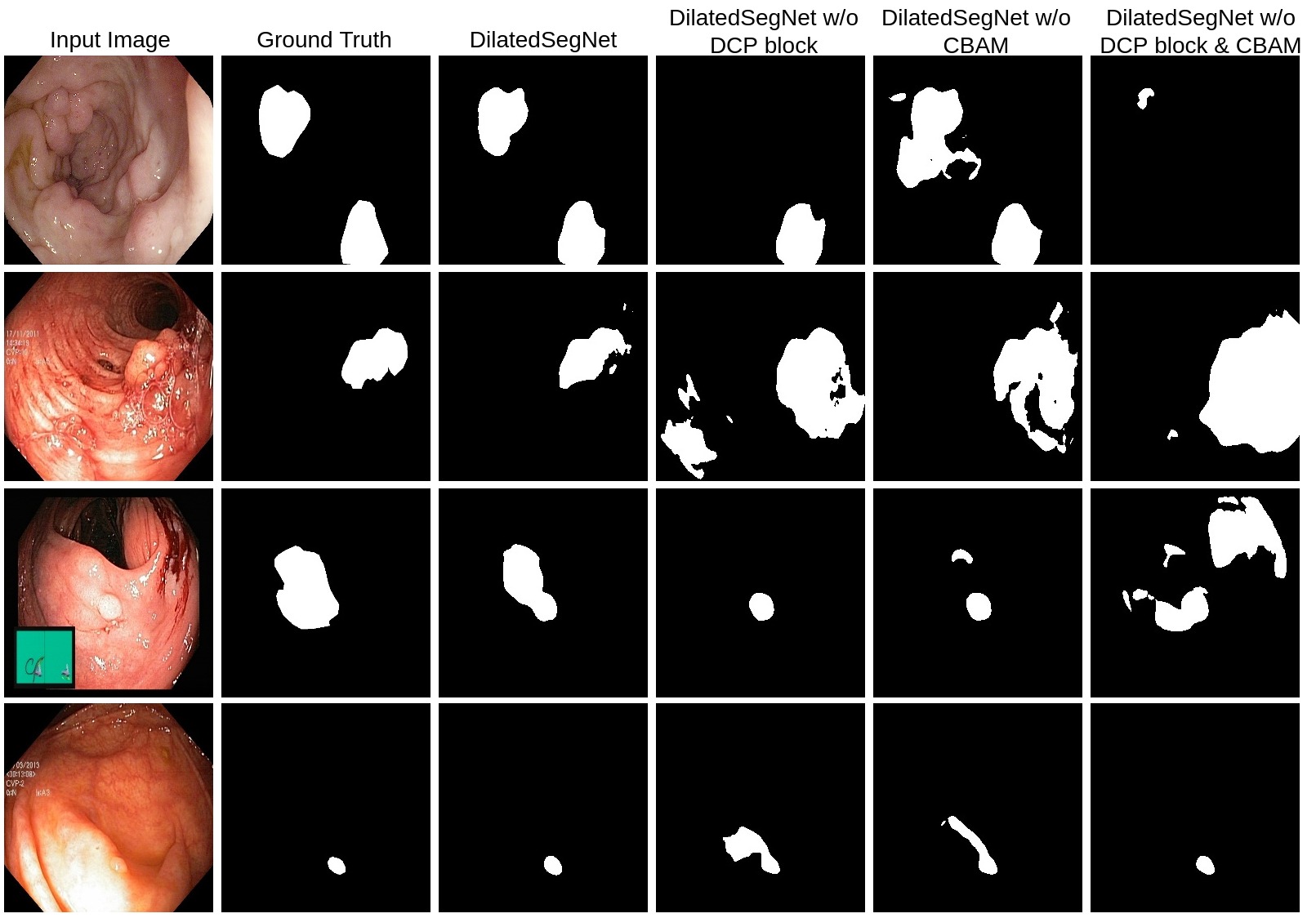}
    \caption{The Figure shows examples of qualitative results comparison of the ablation study from the Kvasir-SEG dataset. The leftmost column shows the input image and the other column next to it shows the ground truth indicating the area covered by polyp and non-polyp. The name of the network used for training during the ablation study is indicated at the top. The qualitative examples show that the proposed network is the best. Eliminating DCP or attention block or both affect the quality of prediction. This is evidenced by the over-segmentation or under-segmentation results produced under the same setting without incorporating the individual or both of the blocks.}
    \label{fig:architecture}
\end{figure*}

Figure~\ref{fig:qualitative} shows the qualitative results of DilatedSegNet and two state-of-the-art networks (i.e., PraNet~\cite{fan2020pranet} and DeepLabv3+~\cite{chen2018encoder}). The qualitative result shows that DilatedSegNet can correctly segment smaller and medium-sized polyps that are commonly missed during routine colonoscopy examinations due to their size. For diminutive polyps, DeepLabv3+ shows over-segmentation and PraNet shows under-segmentation. Similarly, PraNet misses challenging and flat polyps for two cases, and DeepLabv3+ shows under segmentation (for the second example). The visual results comparison shows that DilatedSegNet has a better ability to capture regular and flat polyps. Thus, both the qualitative and quantitative results exhibit the high overall performance of DilatedSegNet. Additionally, we determined the heatmap results of the DilatedSegNet. The heatmap results show the relevance of the individual polyp and non-polyp pixels. The heatmaps can be useful to understand the convolutional neural network and helps towards model interpretability. Here, ``red" and ``yellow" denote the important regions the models learns as polyp, whereas ``blue" color shows that the model considers those regions as less significant areas. 

\subsection{Performance test on completely unseen datasets}

Table~\ref{tab:cross-data} shows the cross-dataset results. In the experimental setting \#1, we train the dataset on Kvasir-SEG and test it on CVC-ClinicDB (a completely unseen dataset). The proposed method obtains a high \ac{DSC} score of 0.8278 and mIoU of 0.7545 and outperforms the best performing DeepLabv3+ by 1.36\% in DSC and 1.57\% in mIoU. Similarly, in setting \#2 when the model is trained on Kvasir-SEG and tested on BKAI-IGH data, the DilatedSegNet surpass the best performing PraNet~\cite{fan2020pranet} and obtains 2.47\% more in DSC and 2.97\% in mIoU.


\section{Ablation study}

\begin{table*} [t!]
\caption{Ablation study of the proposed DilatedSegNet on the Kvasir-SEG~\cite{jha2020kvasir}.}
\centering
\begin{tabular}{@{}l|l|c|c|c|c@{}} 
\toprule
\textbf{No.} &\textbf{Method} &\textbf{DSC} &\textbf{mIoU} & \textbf{Recall} & \textbf{Precision}\\ 
\midrule

\#1 & DilatedSegNet w/o DCP block & 0.8725	&0.8067	&0.8917	&0.9025 \\
\#2 & \shortstack{DilatedSegNet w/o\\ Attention} &0.8832 &0.8135 &0.9076 &0.8966 \\
\#3 & \shortstack{DilatedSegNet w/o DCP block\\ \& Attention (CBAM)} &0.8627 &0.7946 &0.8871 &0.8947 \\
\#4 & DilatedSegNet &\textbf{0.8957}	&\textbf{0.8336}	&\textbf{0.9169}	&\textbf{0.9096} \\

\bottomrule
\end{tabular}
\label{table:ablation-study-table}
\end{table*}
In the Table~\ref{table:ablation-study-table}, we present the results of the ablation study to verify the effectiveness and the importance of each blocks. Here, we test the DilatedSegNet without DCP block (setting \#1), without attention (\# 2), and without DCP block \& attention (\#3).  The proposed architecture has an improvement of 3.3\% in DSC and 3.9\% in mIoU, 2.98\% recall and 1.49\% in precision as compared to the setting \#3. Therefore, we showed that the proposed method had performance improvement with the utilization of DCP and attention block.


\section{Conclusion}
In this work, we proposed the DilatedSegNet architecture that utilizes a dilated convolution pooling (DCP) block and CBAM to accurately segment polyps with high performance and real-time speed, which has never been addressed before. The experimental results on the same dataset testing and completely unseen dataset testing results showed that DilateSegNet achieves a high \ac{DSC} and outperforms the state-of-the-art polyp segmentation models. The design of the architecture was supported by the ablation study. The qualitative, quantitative and heatmap suggest that DilatedSegNet can be a strong benchmark for building early polyp detection in clinics. Additionally, the presented heatmap was effective in discriminating different polyp and non-polyp pixels from the colonoscopy image. In the future, we plan to explore DilatedSegNet with the multi-centre dataset, evaluate its robustness, and explore the results on the federated learning settings. 

 \subsubsection*{Acknowledgement}
 This project is supported by the NIH funding: R01-CA246704 and R01-CA240639.

\bibliographystyle{splncs04}
\bibliography{ref}
\end{document}